\documentclass[12pt]{iopart}
\usepackage{graphicx}%
\usepackage{graphicx}
\relax

\usepackage{color}
\newcommand{\red}[1]{  \textcolor{black}{#1}}

\begin{document}
\bibliographystyle{unsrt}
 \title[Exchange-bias effects in Co/YMnO$_3$ bilayer]{Revealing the origin
 of the vertical hysteresis loop shifts in an exchange biased Co/YMnO$_3$ bilayer}
 \author{J. Barzola-Quiquia\footnote{E-mail: j.barzola@physik.uni-leipzig.de}, A. Lessig, A. Ballestar, C.
 Zandalazini\footnote{Present address: FaMAF-CLCM at University of
Cordoba, Medina Allende S/N, 5000 Cordoba, Argentina.}, G.
Bridoux\footnote{Present address: Institut Catala de
Nanotecnologia (ICN), Universitat Autonoma de Barcelona, E-08193
Bellaterra, Spain.}, F. Bern and P. Esquinazi\footnote{E-mail:
esquin@physik.uni-leipzig.de}}

\address{Division of Superconductivity and Magnetism, Institute for Experimental Physics II, University of
Leipzig, D-04103  Leipzig, Germany}

\begin{abstract}
We have investigated exchange bias effects in bilayers composed by
the antiferromagnetic o-YMnO$_3$ and ferromagnetic Co thin film by
means of SQUID magnetometry, magnetoresistance, anisotropic
magnetoresistance and planar Hall effect. The magnetization and
magnetotransport properties show pronounced asymmetries in the
field and magnetization axes of the field hysteresis loops. Both
exchange bias parameters, the exchange bias field $H_{E}(T)$ as
well as the magnetization shift $M_E(T)$, vanish around the N\'eel
temperature $T_N \simeq 45$~K. We show that the magnetization
shift $M_E(T)$ is also measured by a shift in the anisotropic
magnetoresistance and planar Hall resistance having those a
similar temperature dependence as the one obtained from
magnetization measurements. Because the o-YMnO$_3$ film is highly
insulating, our results demonstrate that the $M_E(T)$ shift
originates at the interface within the ferromagnetic Co layer.
\red{To show that the main results obtained are general and not
because of some special characteristics of the o-YMO$_3$ layer, similar
measurements were done in  Co/CoO micro-wires. The transport and magnetization characterization
of the micro-wires supports the main conclusion that these effects are related to the response
of the ferromagnetic Co layer at the interface.}
\end{abstract}

\pacs{75.60.-d,75.70.Cn}
\submitto{JPCM}
 \maketitle

\bigskip
\section{Introduction}

Exchange bias effects are observed when a ferromagnetic (FM) thin
layer is in atomic contact with an antiferromagnetic (AFM) layer.
These effects are therefore directly related to an exchange
coupling at their interface. Experimentally, they are generally
observed as a shift of the hysteresis loop on the field axis when
the bilayer is cooled from a temperature $T_0$ below the Curie
temperature $T_C$ of the FM layer but above the respective
N{\'e}el temperature ($T_N$) of the AFM layer, to a temperature
$T_1 < T_N$ in the presence of an external magnetic field
$H_{FC}$. \red{While magnetometry studies of the exchange bias
effect and its shift in the field axis $H_E(T)$
\cite{gru00,fra71,rad63} have been deeply performed since its
discovery by Meiklejohn and Bean \cite{mei56} when they
investigated Co particles surrounded by their native
antiferromagnetic oxide, there are a few comparative studies that
used magnetotransport properties to elucidate exchange bias
phenomena \cite{gre02,tri09,shi02,kim03,nic06,nag06}. Specially
rare are the studies using the Hall effect \cite{mcg75}. Recently,
however, the measurement of the exchange bias effects using the
planar Hall effect  has received considerable interest because of
the higher signal-to-noise ratio compared with the
magnetoresistance or spin valve configuration
\cite{ejs04,dam08,hen10}, a feature that might be promising for technology
applications.} On the other hand, the study of metal/oxide
interfaces is a very attracting field that is gaining more
relevance in the condensed matter community \cite{Hwang}.

In the present work, we have investigated the exchange bias
effects in a novel bilayer composed by a polycrystalline
ferromagnetic (FM) Co film and a low temperature antiferromagnetic
(AFM) oxide, namely o-YMnO$_3$ film. A further aim of this work is
to study the exchange bias effects using different properties
namely, SQUID magnetometry, longitudinal magnetoresistance (MR),
anisotropic magnetoresistance (AMR) and the planar Hall effect
(PHE). Besides the usual field shift effect characterized by the
exchange bias field $H_E(T)$, of particular interest was the study
of the magnetization shift $M_E(T)$, an exchange bias effect much
less study in the past probably because of the smaller signal
amplitude and technical difficulties, \red{although recent reports
in special exchange bias systems \cite{zan11,ven12} already showed
that its magnitude can be large}. In this work we show that this
magnetization shift $M_E(T)$ not only can be systematically found
in all measured properties but it has a similar temperature
dependence as $H_E(T)$. The observation of these effects does not
only support the existence of exchange bias in this system and its
influence in both hysteresis loop axes, but it also demonstrates
clearly that the magnetization shift $M_E(T)$ is directly related
to the pinning of magnetic entities (magnetic moments or domain
walls) \cite{fit07,ohlfef} within the FM Co layer at the interface
 and not in the AFM layer, in agreement with magnetization
measurements reported recently in
La$_{0.7}$Sr$_{0.3}$MnO$_3$/o-YMnO$_3$ bilayers \cite{zan11}. This
main conclusion of this study was possible simply because the
transport properties do detect only the FM film and not the AFM
film due to its highly insulating state at the temperatures of the
measurements.

\red{Because the  o-YMO$_3$ layer is not a common AFM but a so
called diluted antiferromagnet in external magnetic field (DAFF),
as will become clear in section~\ref{oymo}, we decided to
characterize both exchange bias effects, i.e. in the field as well
as in the magnetization axis, in the archetype of exchange bias
effects, namely in Co/CoO micro-wires. Our transport as well as
magnetization characterization of the exchange bias effects not
only agree each other but supports the main conclusion we get from
the studies on the Co/o-YMO$_3$ bilayers, namely that the observed
effects originate within the ferromagnetic Co layer, certainly due
to the influence of the AFM layer at the interface. This was
possible to conclude because of the negligible conductance of the
CoO layer in comparison with that of Co. In this case the
magnetoresistance measurements probe the FM Co layer and therefore
may also contribute to the understanding of the FM depth
profile\cite{mor06}. }

\section{Experimental Details}
\label{ed}

We prepared bilayers of FM Co thin films (selected for its weak
anisotropy and small coercivity) covering an AFM orthorhombic
o-YMnO$_3$ (o-YMO) layer grown on (100) SrTiO$_3$ substrates of
area $6 \times 6$~mm$^2$. For the depositions of the o-YMO layer a
KrF excimer laser (wavelength 248~nm, pulse duration 25~ns) was
used. The growth parameters used for o-YMO were 1.7~J/cm$^2$ with
5~Hz repetition rate, $800^\circ$C substrate temperature and
0.10~mbar oxygen pressure during preparation. The thickness of the
o-YMO film for the bilayer discussed in this report was 350~nm.
The Co films were prepared by thermal evaporation under high
vacuum (pressure $P \simeq 10^{-6}~$mbar) using a high purity Co
(99,95\%) precursor material. For the magnetization measurements
we used a superconducting quantum interferometer device (SQUID)
from Quantum Design. The substrate containing the AFM o-YMO layer
was covered completely with 35~nm Co thin film. For the
magneto-transport measurements we used a mask with defined shape
to be able to measure the longitudinal and Hall resistances, see
inset in Fig.\ref{fig6}(b). These measurements were carried out
immediately after the Co deposition and contacting the
corresponding electrodes in a chip carrier.

The epitaxial growth of the o-YMO phase was confirmed by X-ray
diffraction. The electrical transport measurements of the bilayer
(at the temperature of the measurements the electrical transport
is due only to the Co film) were performed with a high resolution
AC bridge (LR700 from Linear Research) in a commercial cryostat in
the temperature range between 5~K and 300~K with magnetic fields
up to 4~T. The resistance of the highly resistive o-YMO film was
measured using a Keithley 6517A electrometer. During the
measurements, the temperature stability was better than 10~mK. A
commercial Hall sensor and a resistance thermometer attached close
to the sample were used to measure the magnetic field and
temperature. The exchange bias effects were studied with the SQUID
and magneto-transport under similar experimental conditions.
Namely, the sample was heated up to 100~K~$> T_N$ and a selected
field was applied parallel to the main area of the sample. Then
the sample was cooled down under this field to the selected
temperature of the measurements. After the temperature was
stabilized, the field was swept in opposite direction and turned
back to the initial field. This process was repeated for each
selected temperature. \red{A systematic study of the influence of
the amplitude of the field $H_{FC}$ used to prepare the field
cooled state showed that the effects of minor loops, as for
example any dependence of the coercive field, are negligible at
$\mu_0 |H_{FC}| \ge 0.3~$T. As was shown already in Ref.~\cite{zan11} for
other bilayer exchange bias systems, the observed exchange bias
effects reported here cannot be attributed to minor loop effects.}

\red{Cobalt thin films ($\simeq 50 \pm 5$~nm thickness) were
deposited by thermal PVD at a base pressure of 10\(^{-7}\)mbar on
a commercial 5 \(\times\) 5 mm$^2$ p-boron doped silicon (100)
substrates,  capped with a 150nm Si\(_{3}\)N\(_{4}\) layer. The
cobalt micro-wires were produced using electron beam lithography
(EBL) in a FEI NanoLab XT 200 Dual Beam microscope with a Raith
ELPHY Plus extension at 10KV acceleration voltage. For the
lithography process a commercial positive-working PMMA (950K)
spin-coated on the substrates with a thickness of about 400\,nm
was used. The samples were prepared in up to three lithography
processes. First the structure for the electrical contacts was
written, which was sputtered in Ar atmosphere with a 5-10nm
Pt/15-25nm Au bilayer. The base pressure of the vacuum chamber was
10\(^{-6}\)mbar, while the working pressure of the Ar sputter gas
was 10\(^{-3}\) mbar. After lift-off and a new resist cover, the
structures for the Co nanowires were written. In this paper we
present the results of two Co/CoO samples,  a single micro-wire of
dimension (length between voltage electrodes $\times$ width)
$8~\mu$m ~$\times$~ $0.7~\mu$m, and an array of 8,500 micro-wires
for the magnetization measurements (length
50$\mu$m, 1$\mu$m width), with a spacing of
20$\mu$m to avoid interaction. The surface of the deposited cobalt
films was analyzed \textit{ex situ} by AFM measurements. The
crystallite size ranges from $\sim$\,25~nm to 40~nm; the surface
roughness is rather small with a peak to peak height variation
smaller than 4~nm. The measurements of the samples were done after
leaving them two months in normal atmosphere for the formation of
a natural, $3 \pm 1~$nm thick oxide film at the Co free surface.
}

\red{The temperature dependence of the  resistivity showed a
linear behavior as one expects for a metal. Below 50~K a quadratic
temperature dependence dominates. The residual resistivity ratio
\(\Gamma = \rho\)(300K)/$\rho$(4K) is rather small ($\Gamma
\approx 1.5$) indicating that the nanowires are highly disordered.
From the residual resistivity $\rho$(4K) $\approx 16~\mu \Omega$cm
we calculated the mean free path $l_{eff} = 6$\,nm $\ldots$ 8\,nm (using the simple Drude formula
$l_{eff}= {m_{e}v_{F}}/{n e^{2}\rho}$\,, with
\(m_{e}=9.1\times 10^{-31}~{\rm kg}, v_{F}=1.5 \times
10^{8}~\mathrm{cm/s}, n= 0.5 \times 10^{23}~\mathrm{cm^{-3}}\)),
a factor of five smaller than the observed grain size.
The temperature dependent resistivity $\rho$(300K)-$\rho$(4K) $
\approx 7.5~\mu \Omega$cm is quite close to the literature values
\cite{CampFert82} giving 10.3$~\mu \Omega$cm in c-direction and
5.5$~\mu \Omega$cm in the perpendicular plane resulting in
7.1$~\mu \Omega$cm for a polycrystalline sample as is our case.
This and the consistency throughout all of our samples indicate
the good quality of the deposited material.}

\section{Results and discussion}\label{res}

\subsection{X-ray characterization}
\begin{figure} %\vskip  9cm  \special{eps:  F:/PCTeXv4/figure1.EPS  x=8.2cm  y=6.2cm}
\includegraphics[width=1.1\columnwidth]{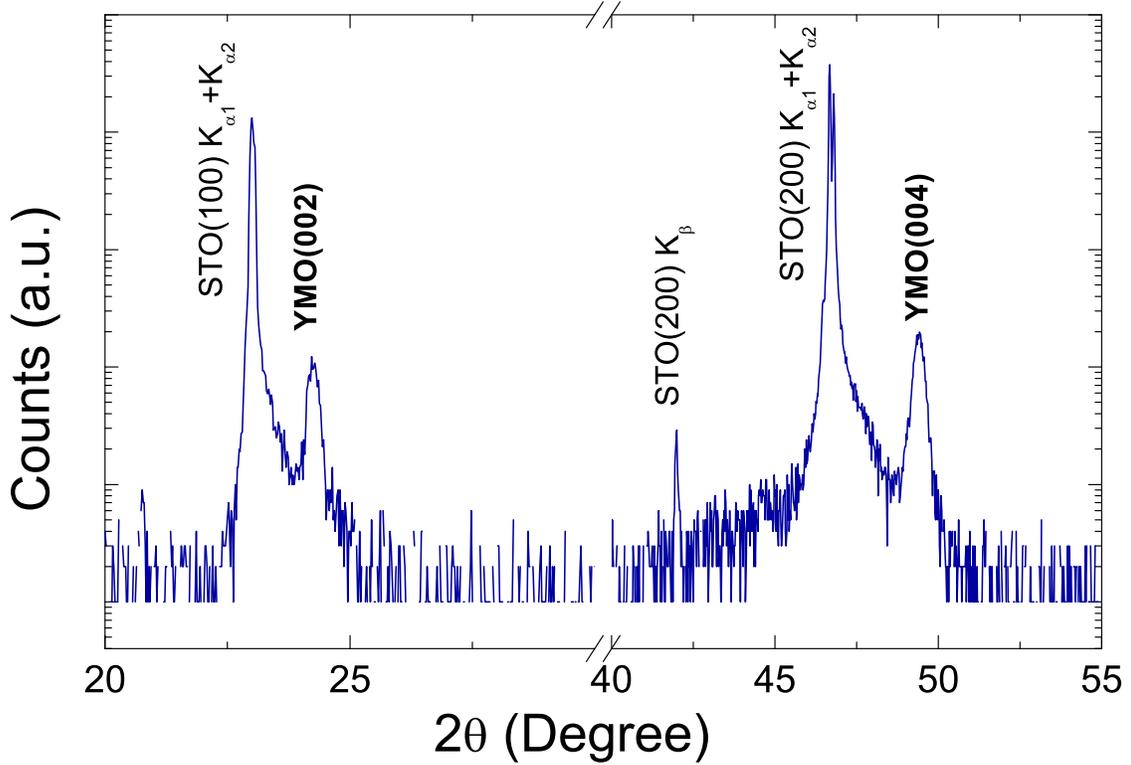}
 \caption{X-ray spectrum of the single o-YMO AFM layer deposited on
 a STO substrate used in our experiments. The thickness of the
 o-YMO film was 350~nm. } \label{xrd}
 \end{figure}

The manganite of yttrium (bulk) usually has a hexagonal structure
\cite{yak63} but under high pressure and temperature it is
transformed in the metastable orthorhombic phase
\cite{wai67,woo73}. This energetically unfavorable phase in
standard conditions can be reached in thin films  if it grows
pseudomorphic with the substrate \cite{gor02}; in this case it can
present different crystal domain structures and lattice strains
\cite{mar08}. We have prepared stable orthorhombic phase of YMO
(o-YMO) epitaxially grown thin films on oriented (100) SrTiO$_3$
(STO) substrates by pulsed laser deposition, as has been reported
earlier \cite{sal98}. The phase of our prepared YMO samples was
confirmed by X-ray diffraction using Cu-K$_\alpha$ line, see
Fig.~\ref{xrd}. The preferential growth of the (00l) planes is
evident and corresponds to the orthorhombic YMO phase
\cite{sal98,mar06}. According to our experimental resolution our
film is purely orthorhombic but the presence of grain boundaries
\red{or certain texture} should not be discarded. From the
position peaks in the $2\theta$-scan, the out-of-plane lattice
parameter is $c = 7.38~\AA$, in good agreement with literature
data \cite{sal98,mar06}. The peaks corresponding to the STO
substrate are also present, see Fig.~\ref{xrd}.

\subsection{SQUID Magnetometry}
\subsubsection{Single o-YMnO$_3$ layer}
\label{oymo} 
\begin{figure} %\vskip  9cm  \special{eps:  F:/PCTeXv4/figure1.EPS  x=8.2cm  y=6.2cm}
\includegraphics[width=0.9\columnwidth]{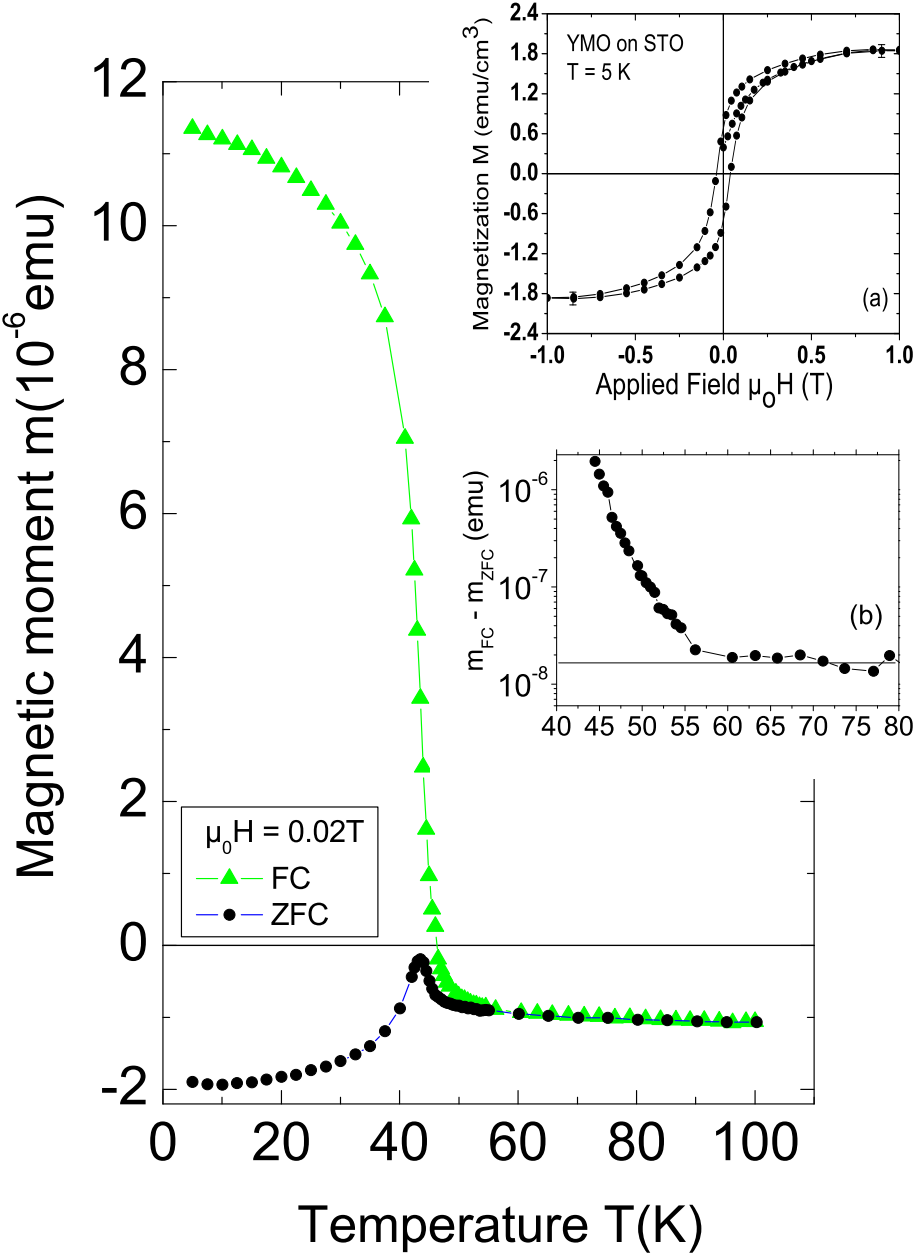}
\caption{Magnetic moment vs. temperature of the o-YMO layer alone
in the FC and ZFC states at a field of 20~mT. The field was
applied parallel to the main area of the sample. In these results
the diamagnetic contribution of the substrate was not subtracted.
\red{The inset (a) shows the field hysteresis of the magnetization
at 5~K. The inset (b) below shows the difference between the two
magnetic moment curves, at FC and ZFC states, of the main panel.}}
\label{mT}
 \end{figure}

According to literature \cite{hsi08,kim06} the o-YMO
phase is AFM with N{\'e}el temperature $T_N = 42 \pm 2~$K. In
Fig.~\ref{mT} we show the magnetic moment of a o-YMO thin film
measured on warming after zero-field cooling (ZFC) and field
cooling (FC) with an applied field $\mu_0H = 20~$mT. From  the ZFC
state we recognize a typical antiferromagnetic behavior. \red{If
we define the characteristic N{\'e}el temperature where the ZFC
magnetic moment shows a maximum we obtain $T_N = 43.5 \pm 0.5$~K
from the data in Fig.~\ref{mT}. If we define $T_N$, however, at
the onset of a significant difference in magnetic moment between
the FC and ZFC states $T_N = 58 \pm 2~$K, see inset (b) in
Fig.~\ref{mT}. It is also known that AFM fluctuations of the order
parameter are established above\cite{dem07} the actual $T_N$,
overestimating in this way the value extracted from SQUID
measurements. Whether the lowest or the upper temperature is the one
where the exchange bias effects vanish, it will depend on the
used experimental method and its sensitivity.}

The behavior observed in the FC state, see Fig.~\ref{mT}, as well
as in previous studies of o-YMO thin films \cite{zan11}, resembles
that of diluted antiferromagnets in external magnetic field
(DAFF). This suggests us to consider the domain state model (DS)
\cite{mil00,kel02,now02,rad03,zab08} to describe the physical
behavior of o-YMO films. \red{It is well known that DAFF develop
a domain state when cooled below $T_N$ (sometimes showing a
spin-glasslike behavior \cite{mun02}) that leads to a net
magnetization which couples to the external field.} The work in
Ref.~\cite{kel02} has demonstrated the broad range of
applicability of this model, being possible to use it not only for
single crystalline AFM, but also for any deviation of the perfect
crystalline structure of the AFM, such as punctual defects, grain
or twin boundaries. In particular, it has been applied
in the exchanged-bias Co/Co$_{1-y}$O bilayer, where the
Co$_{1-y}$O AFM layer presents crystallite sizes between $25$ and
$35~$nm. Finally, it is important to note that the magnetic moment
at saturation of our YMO film in the FC state is orders of
magnitude smaller than that of  the Co film. \red{The inset (a) in
Fig.~\ref{mT} shows the hysteresis loop of the magnetization of
the YMO single layer at 5~K, a loop characteristic of an AFM, being
symmetric and without any vertical or horizontal shift. The
magnetic moment at saturation and at 5~K is equal to 23~$\mu$emu.
This means that the magnetic moment ratio at saturation between
the FM and AFM layers for the bilayer sample is $\sim 10^2$. The
tendency to saturation of $m(T\rightarrow 0)$ at a field of only
0.02~T shown in Fig.~\ref{mT} is also typical for conventional
DAFF materials.}

\red{Taking into account the results on the
Co/CoO micro-wires we present below in section~\ref{cocoo}
we believe that a further discussion on whether our thin o-YMnO$_3$ layer
behaves or not as a typical DAFF or whether this material
is what we have at the interface with the layer Co, does not
appear to us relevant, as will become clear after the presentation and
discussion of the main results.}

\subsubsection{Bilayer Co/o-YMnO$_3$}

\begin{figure} %\vskip  9cm  \special{eps:  F:/PCTeXv4/figure1.EPS  x=8.2cm  y=6.2cm}
\includegraphics[width=0.9\columnwidth]{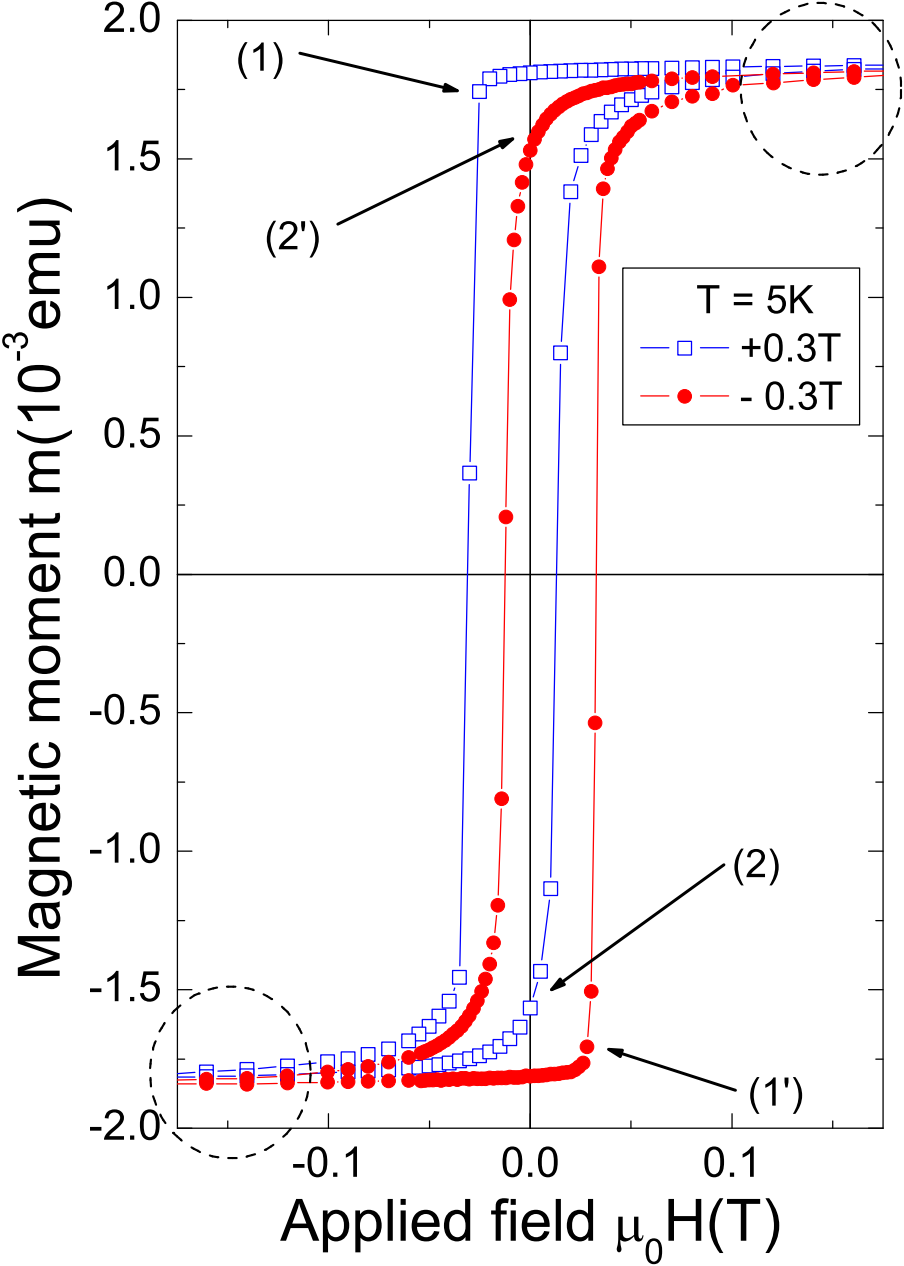}
\caption{Hysteresis loops of the Co/o-YMO bilayer cooled to $T =
5~$K at two fields $\mu_0 H_{FC} = \pm 0.3~$T. The two dashed
circles indicate the regions where the vertical shift, the $M_E$
effect, can be quantitatively characterized.} \label{fig3}
 \end{figure}

It is well known that the exchange bias corresponds to an exchange
field $(H_E)$ transferred from the surface of the AFM into the FM,
but its direction depends on the type of exchange coupling at the
interface between the FM and AFM layers \cite{zab08,nog99}. Upon
the magnetic characteristics of the layers material used (and in
some cases also on the intensity of the applied field
\cite{nog96}) it is possible to found negative (usual case) as
well as positive exchange bias \cite{kiw00} if the shift of the
hysteresis loop along the field axis is in the opposite or same
direction as the applied field, respectively. On the other hand, a
vertical (magnetization) shift is observed if a number of frozen
magnetic moments  in one of the layers at the interface remains
uncompensated due to the proximity to the other layer. Their
orientation can be parallel or antiparallel with respect to the FM
layer  generating in this way a shift upwards of the hysteresis
loops in case of a direct exchange or a shift downward in case of
an indirect exchange mechanism \cite{zab08,nogPRB}.

In order to corroborate the existence of the above described
effects in our bilayer we have measured it by cooling with
positive and negative magnetic fields ($H_{FC} = \pm 0.3~$T). The
results obtained at $T = 5~$K are shown in Fig.~\ref{fig3}. The
shift in the magnetic field axis is clearly observed. We use the
phenomenological definition for the strength of this effect
characterized by the ``exchange bias field'' defined as $H_E =
(H_C^+ + H_C^-)/2$, where $H_C^+$ and $H_C^-$ are the coercive
fields at positive and negative fields. In Fig.~\ref{fig3} we
observe that the expected hysteresis shift depends on the sign of
the $H_{FC}$. Within experimental resolution the absolute value of
$|H_E(5$K$)| \simeq 9~$mT is similar for both directions of
$H_{FC}$.

The hysteresis loops show an asymmetry  near the coercive fields
where a change in the magnetization direction occurs. When the
field is swept in the opposite direction as the one of $H_{\rm
FC}$, the magnetization remains constant up to the coercive field
(positions (1) and (1') in Fig.~\ref{fig3}) before a sudden change
in the magnetic moment occurs. After reaching the saturation and
reversing the field  to the initial value, the magnetization curve
remains nearly constant till the other coercive field is reached
where the magnetic moment shows a rounded edge (arrows 2 and 2').
\red{This asymmetric behavior in the magnetization is already
known and it may have two different origins in an exchange bias
system. One of them is related to a stable interface without
training effect producing asymmetries explained in terms of a
competition between magnetic field orientation, ferromagnetic and
exchange-bias anisotropies \cite{cam05}. The other one is
intimately related to irreversible changes of the domain
structure during the magnetization reversal, i.e. a training
effect found in CoO/Co bilayers \cite{gie02,hof04,bre05}, for
example. It is directly linked to a change in the magnetization
reversal mechanism as  a consequence of domain wall nucleation,
domain wall propagation and rotation of the magnetization of the
domains \cite{rad03}, i.e. an origin directly related to the
ferromagnetic layer influenced by the interface. Although we do
not present a direct study of training effects, the similarities
of the observed asymmetries in the field loops in the
magnetoresistance, a property directly related to the Co layer
only, see Fig.~\ref{fig7}(a),  suggest that the observed asymmetry
features have the same origin as in Co/CoO
 bilayers \cite{gie02,hof04,bre05}.}

A detailed inspection of the data in Fig.~\ref{fig3} reveals a
small but systematic shift in the magnetic moment axis. This shift
(at the regions within the dashed circles in Fig.~\ref{fig3})
depends on the field direction used during the cooling process. In
order to quantify the  $M_E$ effect we define the magnetic moment
shift as $m_{\rm shift} = (m_s^+ + m_s^-)/2$, where $m_s^+$ and
$m_s^-$ are the saturation moments at  certain  positive  and
negative fixed fields with the same absolute value. Our
experimental setup for magnetic moment measurements uses a home
developed sample holder \cite{barzola1}, which allows a
reproducibility and sensitivity of $\simeq 3 \times 10^{-7}$~emu
at a field of 1~T. At the field of $\pm 0.3~$T the reproducibility
is of the order of $1 \times 10^{-7}$~emu. From the data in
Fig.~\ref{fig3} we obtain at $ T = 5~$K a $m_{\rm shift} = (9 \pm
0.5) \mu$emu at $\mu_0 H_{FC} = 0.3~$T (see Fig.~\ref{fig5}),
nearly two orders of magnitude larger than our resolution and
reproducibility limits. \red{Note that this magnetic moment shift
is huge, compared with the total magnetic moment at saturation of
the AFM layer alone. In fact and taking the magnetization of the
YMO layer alone, see inset (a) in Fig.~\ref{mT},  the total
magnetic moment at saturation of the YMO layer is $23~\mu$emu. The
observed $m_{\rm shift} = 9~\mu$emu would mean that about 135~nm
YMO thick layer should contribute to the shift, a value obviously
much larger than the expected exchange bias interface thickness.
On the other hand magnetization loops of the YMO alone at
different cooling fields $H_{FC}$ show hysteresis loops without
any vertical or horizontal shifts. Therefore, the origin of the
$M_E$ effect should not be in the YMO but in the Co layer at the
interface, as the transport measurements also indicate.}

The $M_E$ effect in our bilayer can be interpreted as follows.
After field cooling in an external field and due to the presence
of the interfacial exchange field of the ferromagnetic layer, the
AFM layer develops a frozen domain state that controls the
exchange bias and the magnetic domains within the FM/AFM interface
layer producing an irreversible surplus of magnetization at the FM
side, as the experimental results indicate and in agreement with
observations on La$_{0.7}$Sr$_{0.3}$MnO$_3$/o-YMnO$_3$ and
La$_{0.67}$Ca$_{0.33}$MnO$_3$/o-YMnO$_3$ bilayers \cite{zan11}.
The formation and number of the domains within the FM Co layer
that take part in the exchange bias coupling with the AFM layer
can be enhanced leading to an increase of $M_E$. Therefore, the
measured $m_{\rm shift}$ can be easily explained by pinned moments
(or  magnetic domains by domain wall pinning) within the FM layer
at the interface, i.e. a pinned thickness of the order of 0.2~nm
Co layer would be enough to produce the observed $m_{\rm shift}$.

\begin{figure} %\vskip  9cm  \special{eps:  F:/PCTeXv4/figure1.EPS  x=8.2cm  y=6.2cm}
\includegraphics[width=0.9\columnwidth]{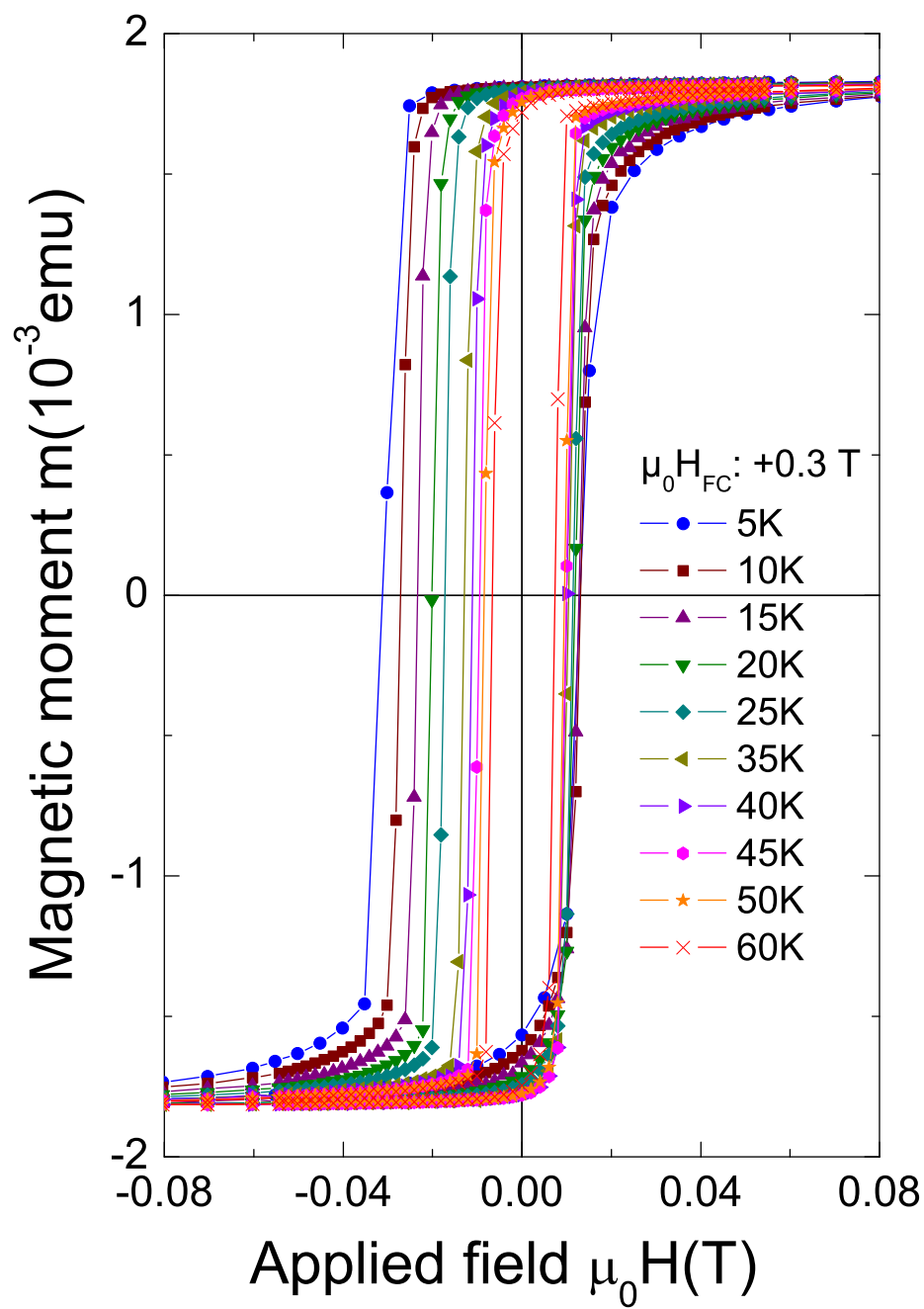}
\caption{Hysteresis loops from SQUID measurements at different
constant temperatures showing the exchange bias shift after
cooling the sample with a positive magnetic field.} \label{fig4}
 \end{figure}

\begin{figure} %\vskip  9cm  \special{eps:  F:/PCTeXv4/figure1.EPS  x=8.2cm  y=6.2cm}
\includegraphics[width=0.9\columnwidth]{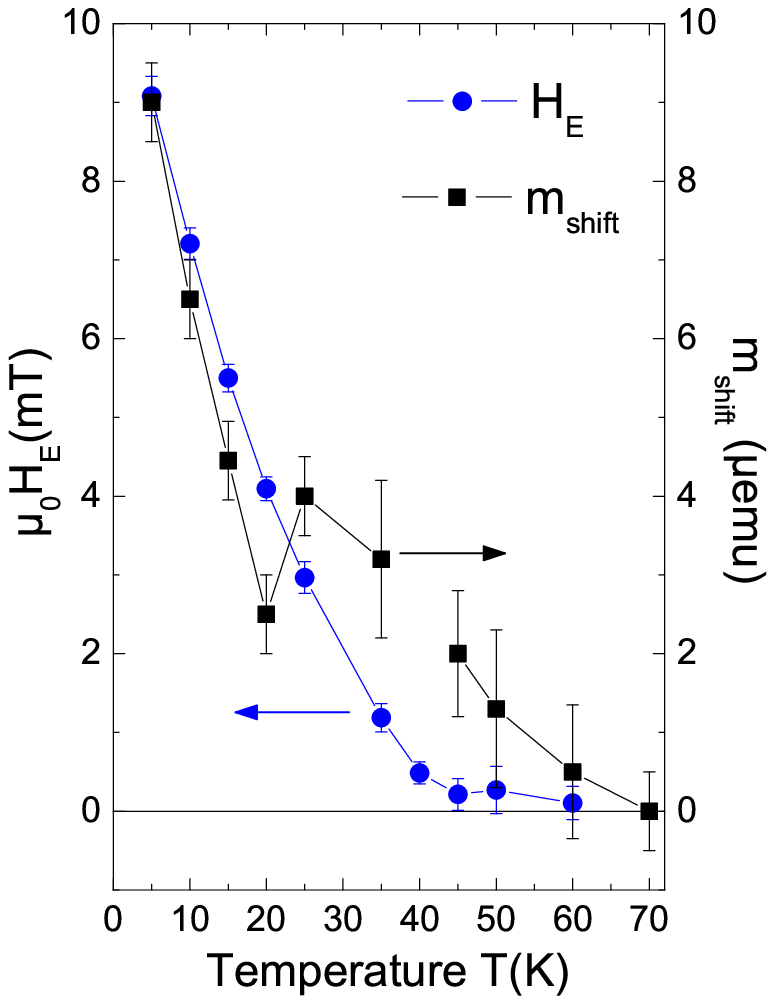}
\caption{Temperature dependence of the exchange bias shift
parameters obtained from the SQUID measurements after cooling the
sample with a positive magnetic field of $+0.3$~T.} \label{fig5}
 \end{figure}

Figure~\ref{fig4} shows all the performed SQUID measurements at
different temperatures when the sample was cooled with a positive
$H_{\rm FC}$. The exchange bias field $H_E$ as well as the vertical
shift $m_{\rm shift}$ in the magnetic moment obtained from the data
in Fig.~\ref{fig4} are shown as a function of temperature in
Fig.~\ref{fig5}. These two quantities vanish between 45~K and 60~K,
near the onset of the AFM transition, see Fig.~\ref{mT}, indicating
that the measured effects should be a result of  the exchange bias
effect. At $T < 30~$K the exchange bias field follows $H_E(T)
\propto (1-T/T_N^\star)$, where $T_N^\star \simeq 32~$K is a kind of
blocking temperature in general lower than the N\'eel temperature
\cite{gru00}. This linear dependence in $T$ has also been observed
in bilayer systems like FeO/Fe, CoO/Co and NiO/Ni \cite{gru00} and
theoretically predicted assuming a cubic anisotropy for the AFM
material \cite{mal88}. Note that the first sudden change in the
magnetization (at (1) in Fig.~\ref{fig3}) begins to be rounded when
the temperature approaches $T_N$, see Fig.~\ref{fig4},  as a
consequence of the weakening of the exchange coupling when the
temperature increases. We note that both parameters $H_E(T)$ and
$m_{\rm shift}(T)$ show a similar -- but not identical --
temperature dependence, see Fig.~\ref{fig5}. In particular there is
a clear change of slopes at $\sim 30$~K and $\sim 20$~K in $m_{\rm
shift}(T)$, which is not observed in $H_E(T)$ within the resolution
of the experiment. This anomaly in the temperature dependence is
related very probably to the ferroelectric transition of o-YMO
observed at $T \sim 30~$K \cite{hsi08,kim06}.

\subsection{Magneto-transport measurements}

\begin{figure} %\vskip  9cm  \special{eps:  F:/PCTeXv4/figure1.EPS  x=8.2cm  y=6.2cm}
\includegraphics[width=1.1\columnwidth]{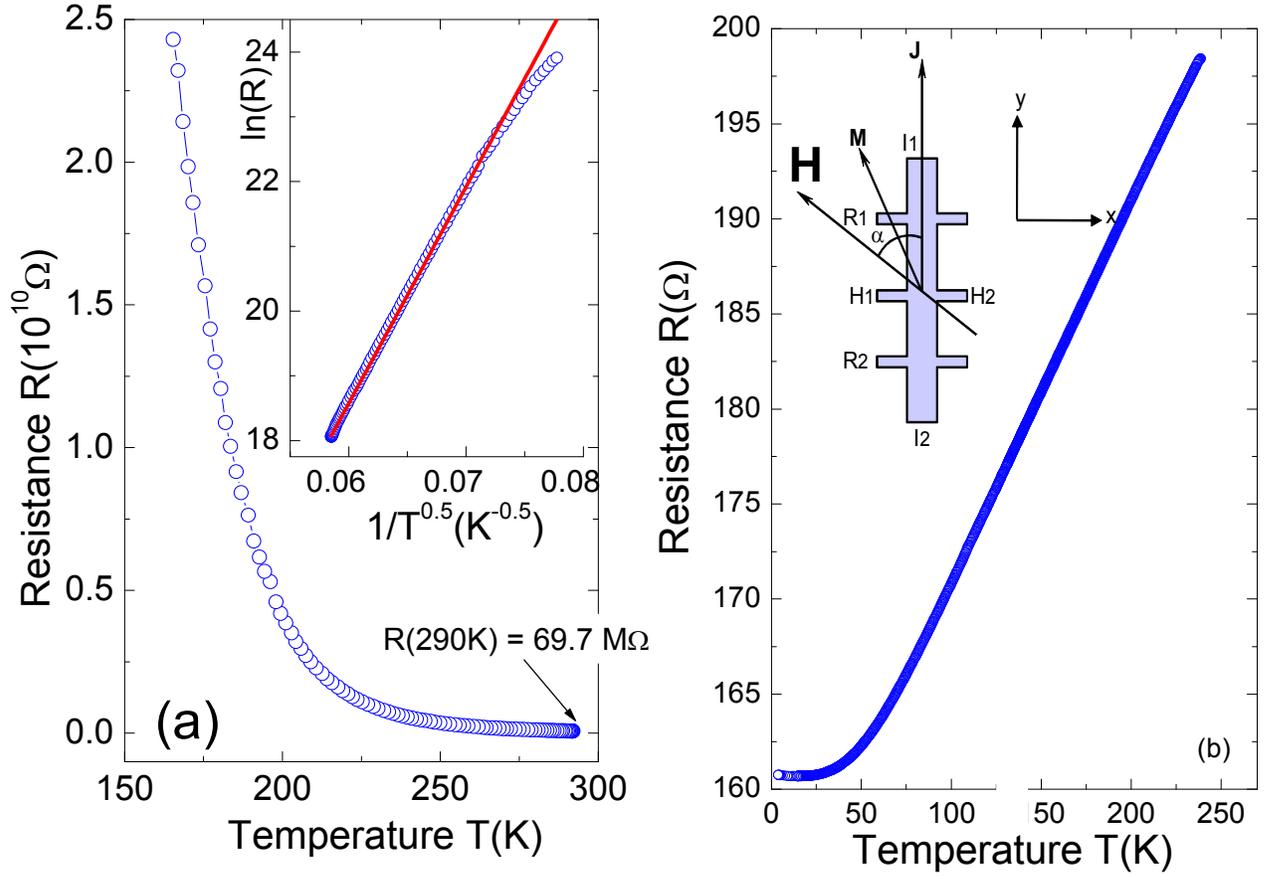}
\caption{Temperature dependence of the resistance  for (a) the YMO
film alone, and (b) the Co film deposited on the top of the YMO.
The inset in (a) shows the same data for the YMO layer but in a
semilogarithmic scale vs. $T^{-0.5}$ and the inset in (b) shows
the sample geometry and electrodes used for the measurements of
the bilayer. The arrows show the field $H$ applied at an angle
$\alpha$ from the main (current) axis and an arbitrary
magnetization direction $M$ out of the main axis and on the sample
main plain.} \label{fig6}
 \end{figure}

In our bilayer the o-YMO layer deposited on  STO serves as a
dielectric substrate allowing us to measure the transport properties
of the structured Co film deposited on the top. In order to warranty
that the resistance contribution of the o-YMO layer is negligible
compared to the Co layer, we have characterized similar YMO films
using an electrometer with an upper resistance limit of $\sim
10^{10}~\Omega$. The corresponding results are plotted in
Fig.~\ref{fig6}(a). The resistance of the o-YMO layer of similar
thickness and area as the one used for the bilayer, shows a
resistance of $\simeq 70~$M$\Omega$ at 290~K, in contrast to the
resistance $\sim 200~\Omega$  of the Co layer, see
Fig.~\ref{fig6}(b). The increase in the resistance of the o-YMO
layer decreasing temperature is huge and for $ T < 160~$K the
resistance is larger than the maximum  we can measure. The
resistance of the  o-YMO film follows a variable range hopping (VRH)
behavior given by the Efros-Shklovskii VRH dependence \cite{efr75},
as the inset in Fig.~\ref{fig6}(a) indicates. The small deviation at
low temperatures is an artifact due to the upper limit of the
electrometer. Consequently, especially for temperatures $T < 100~$K
 the o-YMO layer behaves as an insulator in
comparison to the resistance of the Co layer, giving us the
possibility to explore the exchange-bias effects using
magneto-transport properties that provide us direct
information about the magnetic behavior of the FM layer and the
influence of the exchange coupling of the AFM layer on it.

%In particular, the measurement of the PHE  allows us to obtain the
%exchange bias effect that correlates to the $m_{\rm shift}$.

Figure~\ref{fig6}(b) shows the temperature dependence of the
resistance of the Co film deposited on the top of the o-YMO film. As
expected, the observed behavior is typical for polycrystalline
metallic Co in agreement with reported data \cite{gil05}. The inset
in this figure shows a sketch of the patterning of the Co sample
used in this measurement and the magnetic field and electrical
current directions.

\begin{figure} %\vskip  9cm  \special{eps:  F:/PCTeXv4/figure1.EPS  x=8.2cm  y=6.2cm}
\includegraphics[width=0.8\columnwidth]{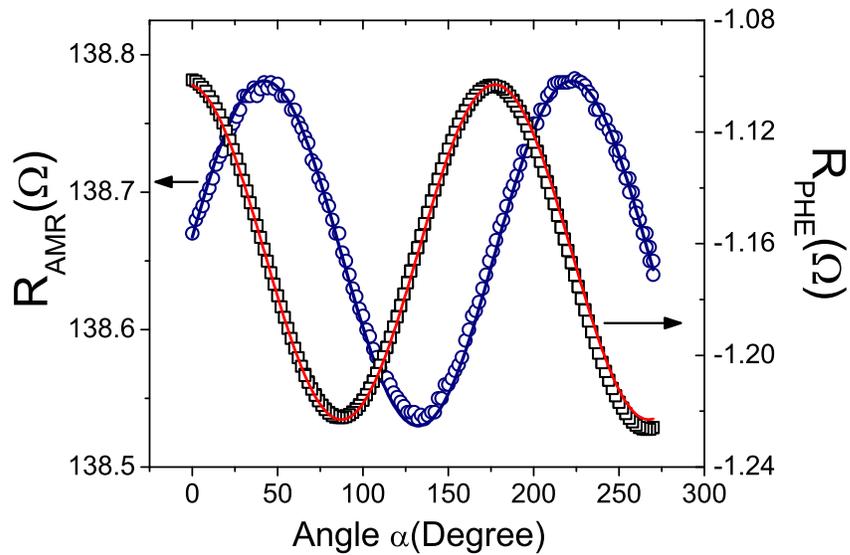}
\caption{Anisotropic magnetoresistance (AMR) and planar Hall
effect (PHE) measured at room temperature using a field of $\mu_0
H = +0.25~$T. Due to the alignment of the sample inside the sample
holder at an angle $\alpha = 43^\circ$ the magnetic field is
parallel to the main samples axis, i.e. parallel to the current
direction. The continuous lines are the fits to Eqs.~(1) and (2)
with the same parameter $R_{\parallel}-R_{\perp} = 0.25~\Omega$
and $R_{\perp} = 138.53~\Omega$. } \label{figV1}
 \end{figure}

Our experimental setup allows us to perform angular dependent
measurements changing the angle  $\alpha$, see Fig.~\ref{fig6}(b),
from  $0 ^\circ$ to $270^\circ$. It is known that
magneto-electrical transport measurements on conducting
ferromagnets can provide information about their magnetic
properties, as, e.g., the transition temperature, the easy axis
direction, the coercivity, etc.  We have measured the so called
anisotropic magnetoresistance (AMR) and the planar Hall effect
(PHE). The experimental conditions for such measurements are
similar to the classical magnetoresistance (MR) and Hall effect,
but the  magnetic field \textbf{H} is applied parallel to the
sample main plane formed by  \emph{x} and \emph{y} (see
Fig.~\ref{fig6}(b)). For this field direction the AMR is the
resistance measured between electrode R1 and R2 of
Fig.~\ref{fig6}(b)), and the PHE   is the voltage measured between
electrodes H1 and H2 of Fig.~\ref{fig6}(b). Assuming a
single-domain model, the AMR as a function of the angle is given
by

\begin{equation}
\label{eqone} R_{AMR}(\alpha)= R_{\perp} +
\left(R_{\perp}-R_{\parallel}\right) \cos^2\alpha\,,
\end{equation}
and the PHE is

\begin{equation}
\label{eqdos} R_{PHE}(\alpha)=
\left(R_{\parallel}-R_{\perp}\right) \cos\alpha\sin\alpha\,,
\end{equation}
where $R_{\parallel}$ and $R_{\perp}$ are the resistances when the
magnetization is parallel and perpendicular to the current,
respectively. The difference between $R_{\parallel}$ and $R_{\perp}$
is the origin for the anisotropic magnetoresistance and the planar
Hall effect. Figure~\ref{figV1} shows the AMR and PHE  at
room temperature as a function of the angle $\alpha$. The good fits
using the above equations confirm the single-domain model
assumption at the applied field. Due to the large  value of $R_{\perp}$,
one can realize that the relative change in the PHE
is larger than the one of  AMR, making the first one a more sensitive
tool to study magnetic effects produced by the
exchange bias.

\begin{figure} %\vskip  9cm  \special{eps:  F:/PCTeXv4/figure1.EPS  x=8.2cm  y=6.2cm}
\includegraphics[width=0.9\columnwidth]{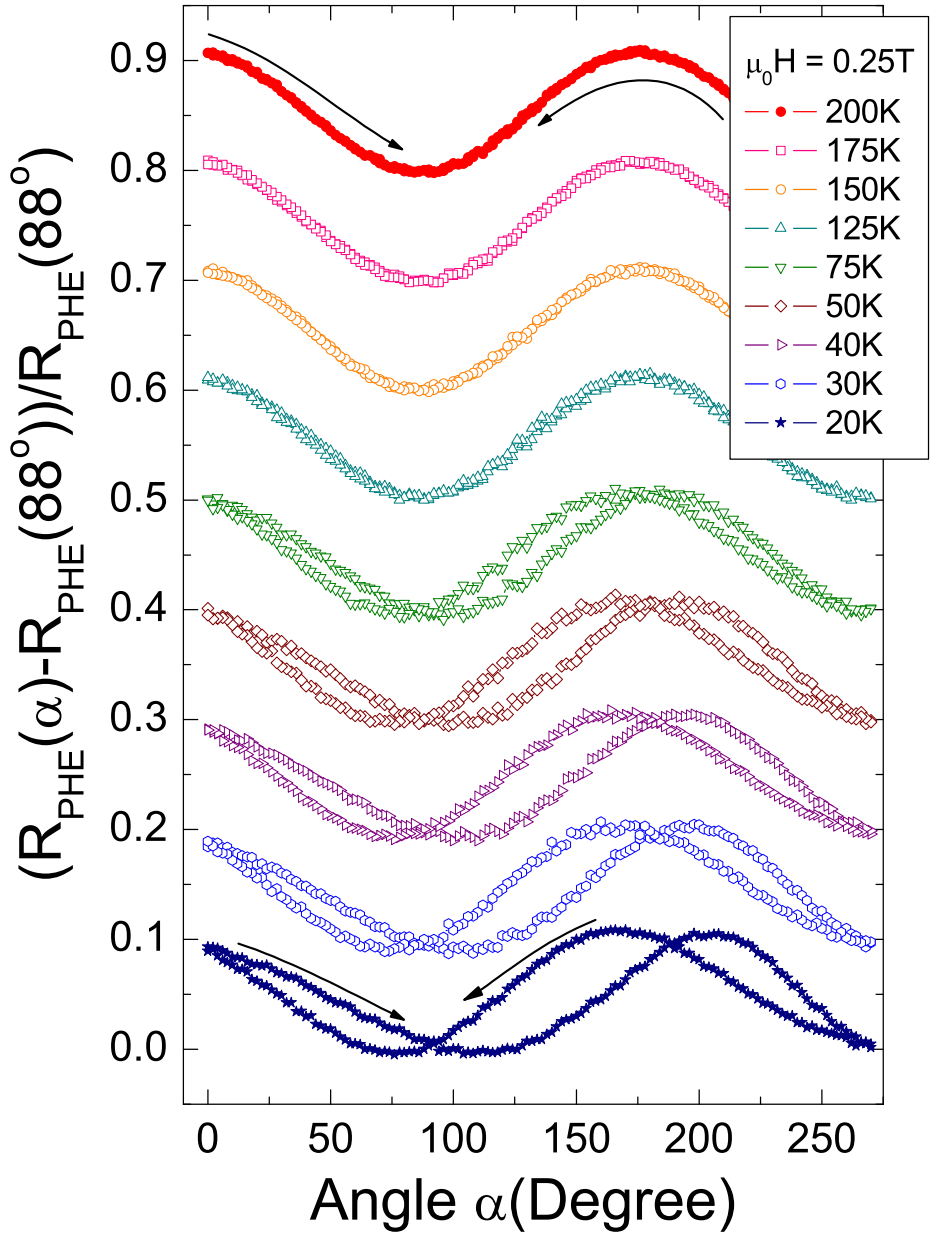}
\caption{PHE measurements in cycle modus at different constant
temperatures and at constant applied field  of $\mu_0 H = 0.25~$T.
The curves are normalized at their corresponding $R_{PHE}$ values
at $\alpha = 88^\circ$. Note that the initial and final values are
the same within experimental uncertainty. The measurements were
done starting always from the ZFC state at each temperature.}
\label{figV3}
 \end{figure}

\begin{figure} %\vskip  9cm  \special{eps:  F:/PCTeXv4/figure1.EPS  x=8.2cm  y=6.2cm}
\includegraphics[width=0.7\columnwidth]{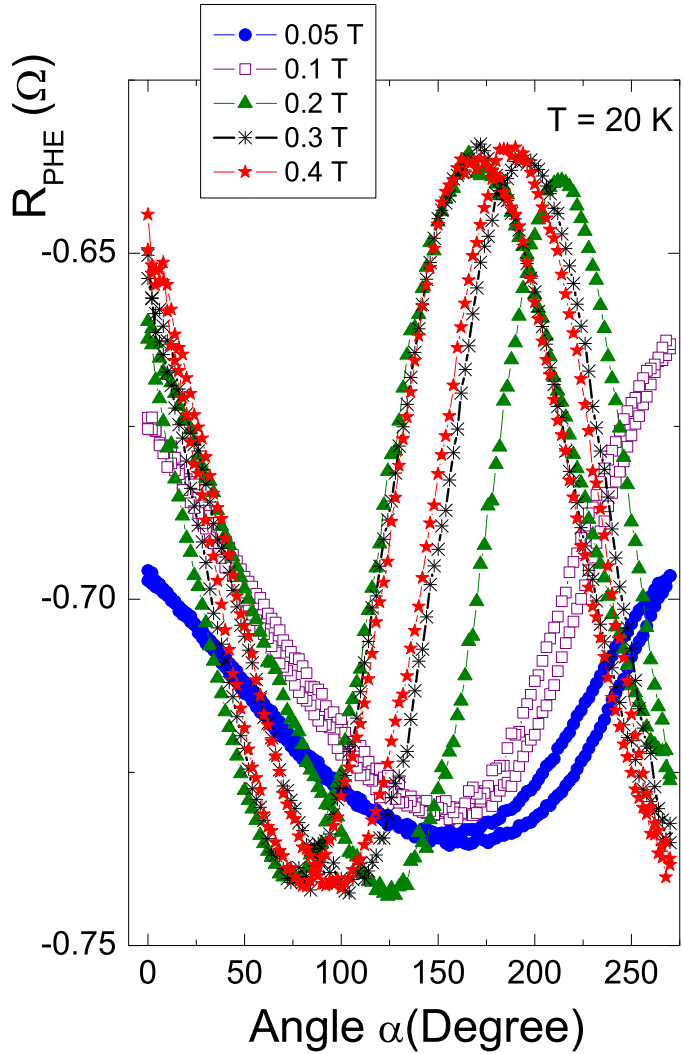}
\caption{PHE measurements in the ZFC state and at $T=20~$K at
different constant applied fields.} \label{figV4}
 \end{figure}

We have performed the PHE measurements in a ZFC state at different
constant temperatures, i.e. above and below $T_N$, and the angular
dependent measurements cycling from $\alpha = 0^\circ$ to $\alpha
= 270^\circ$ and back to $\alpha = 0^\circ$. At temperatures above
$100~$K the results follow Eq.~(2), see Fig.~\ref{figV3}, while
below $100~$K a hysteresis is observed although there is no
difference between the initial and final $R_{PHE}$ values within
experimental resolution. This reversibility of the initial and
final $R_{PHE}$ values after cycling the field remains also at
different applied fields as Fig.~\ref{figV4} shows. From these
results we conclude that independently of the applied field, starting always from
a ZFC state, there
is no irreversibility between the initial and final values of the
PHE after cycling the angle. The hysteresis observed at angles
$\alpha \neq 0 ^\circ$ is a natural consequence of the cobalt
ferromagnetic hysteresis.

\begin{figure} %\vskip  9cm  \special{eps:  F:/PCTeXv4/figure1.EPS  x=8.2cm  y=6.2cm}
\includegraphics[width=0.7\columnwidth]{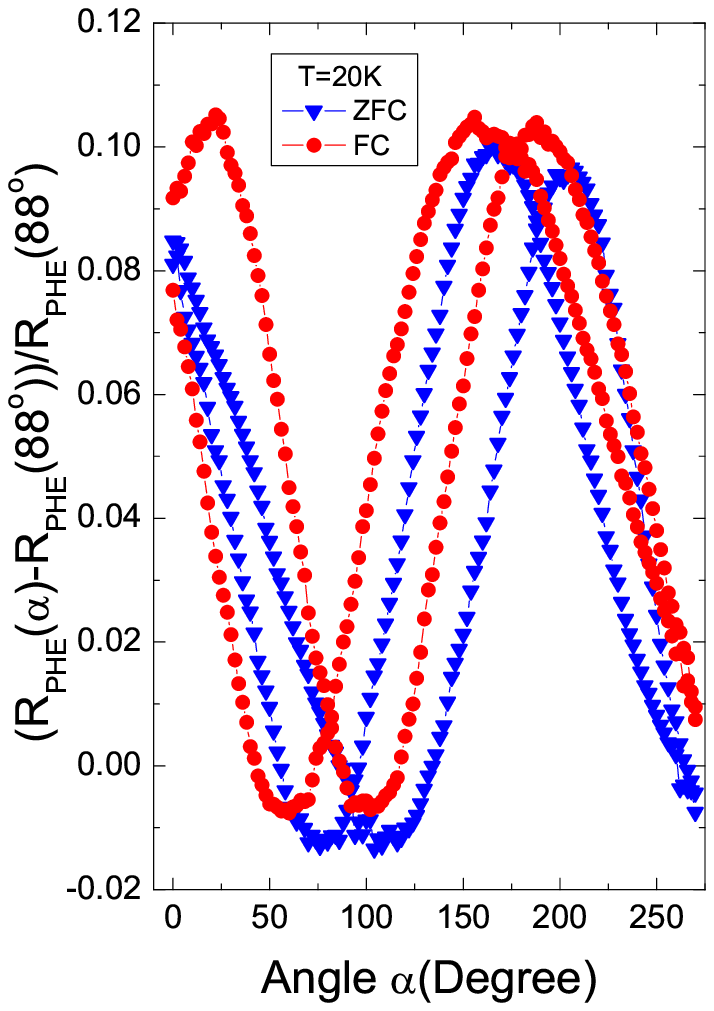}
\caption{PHE measurements (normalized at their corresponding
values for $\alpha = 88^\circ$) as a function of the angle
$\alpha$ at a constant applied field $\mu_0 H = 0.25~$T. The
measurements were done at $T= 20~$K in a ZFC and FC states.}
\label{figV5}
 \end{figure}

\begin{figure} %\vskip  9cm  \special{eps:  F:/PCTeXv4/figure1.EPS  x=8.2cm  y=6.2cm}
\includegraphics[width=0.8\columnwidth]{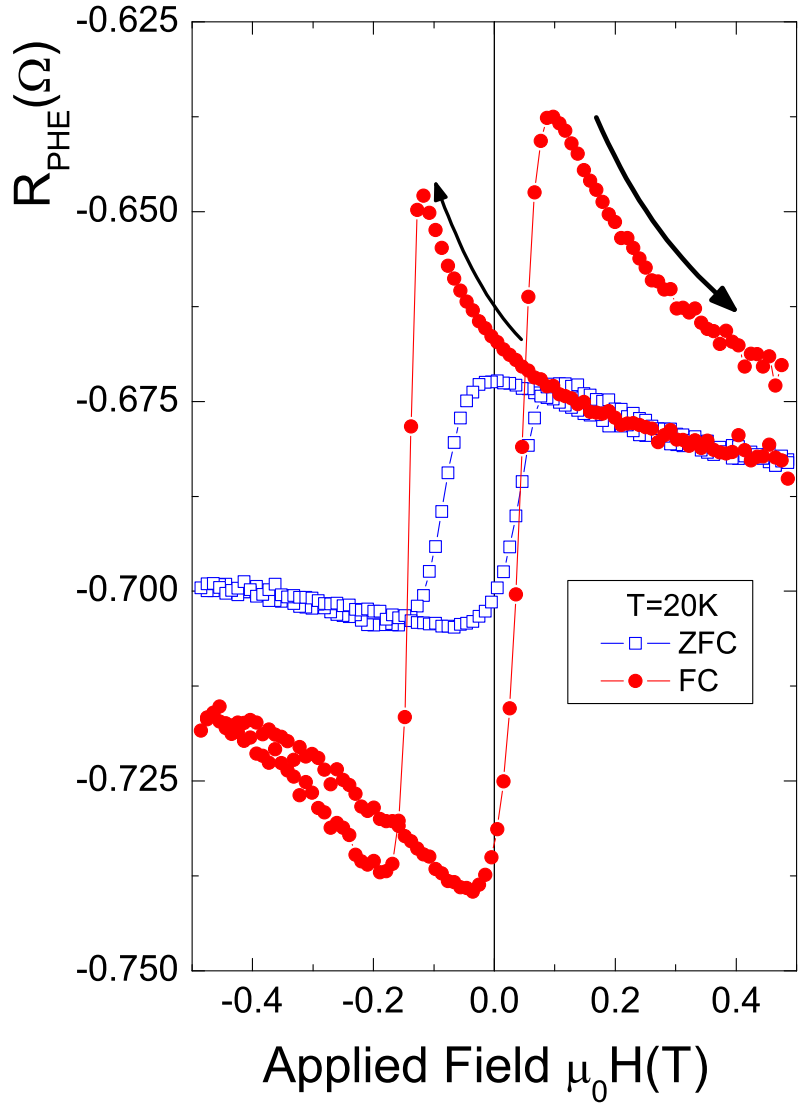}
\caption{PHE measurement at fixed angle corresponding to the
situation where the magnetic field \textbf{B} and current
direction are parallel. The measurements were done at $T= 20~$K in
the ZFC and FC ($\mu_0 H_{\rm FC} = 1~$T) states.} \label{figV6}
 \end{figure}

Figure~\ref{figV5} shows PHE measurements in the FC state where it
becomes evident that after a cycle the PHE resistance at
the end of the cycle is not the same, i.e. a clear PHE-shift
appears at the end of the loop in clear contrast to the ZFC
measurements. On the other hand, the hysteresis observed at angles
$\alpha \neq 0 ^\circ$ in ZFC is also present in the FC state.
Consequently, this hysteretic feature, the PHE-shift at the end of
cycle, should be directly related to an exchange bias effect.
After this experimental observation we have fixed the angle
$\alpha$ such that $\textbf{M}$ was parallel to the main axis of
the sample and performed a field dependent measurement of the PHE
at FC and ZFC states, see Fig.~\ref{figV6}. We found open and
close hysteretic loops for the FC and ZFC state, respectively, in
good agreement with the results of Fig.~\ref{figV5}. For a fixed
magnetic field value, see Fig.~\ref{figV6}, the PHE resistance
increases in the FC state indicating that it is highly sensitive
to the exchange bias effects due to pinned magnetic moments or
domains at the interface. These results are compatible with a
scenario where an interfacial domain structure is present in the
ferromagnetic Co layer \cite{rad03,bre05}.

\begin{figure} %\vskip  9cm  \special{eps:  F:/PCTeXv4/figure1.EPS  x=8.2cm  y=6.2cm}
\includegraphics[width=0.9\columnwidth]{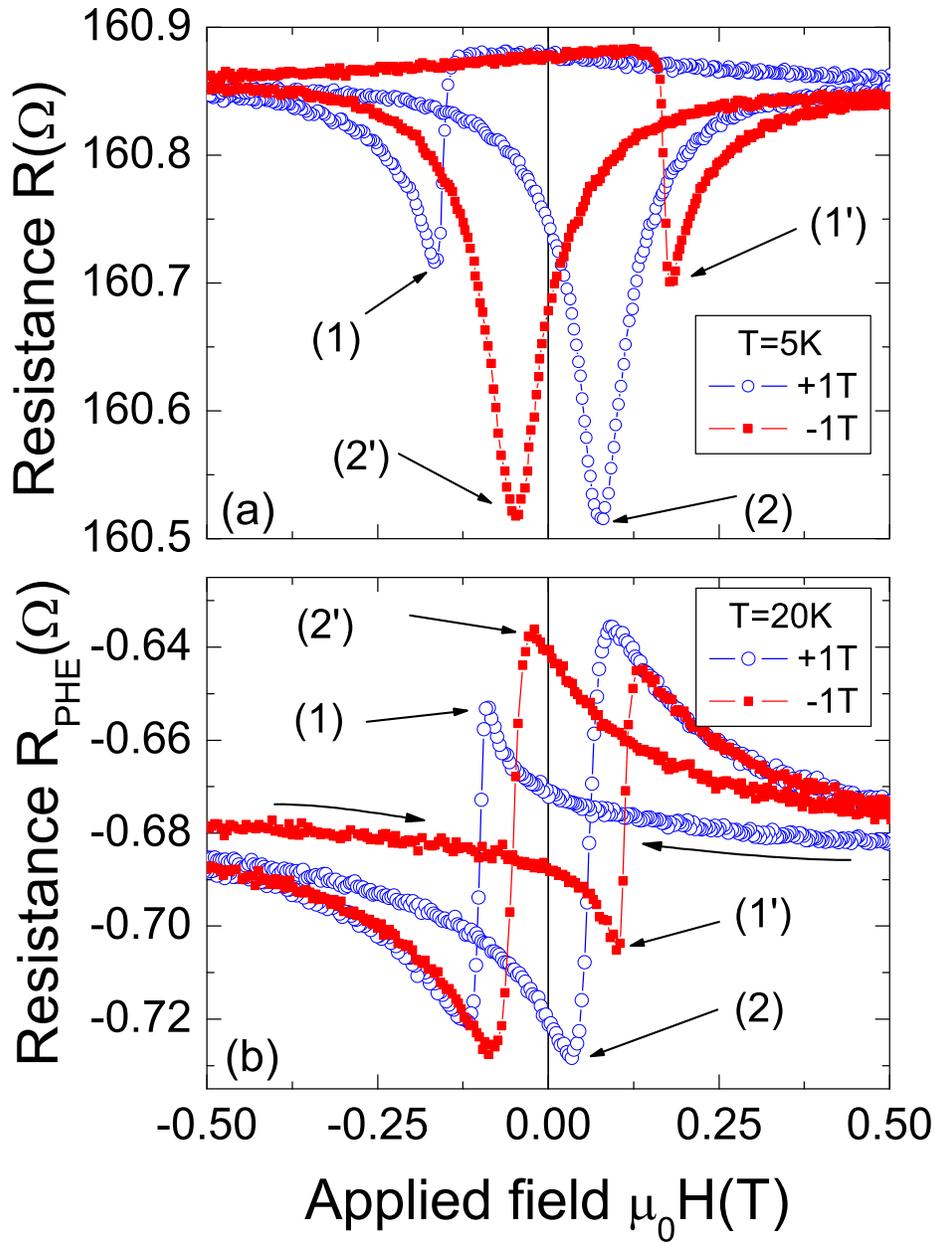}
\caption{(a) Magnetoresistance and (b) the planar Hall effect
measured at 5~K and 20~K, respectively, using positive and
negative cooling field $|\mu_0 H_{\rm FC}| = 1~$T. The numbers in
brackets indicate similar features as observed in magnetization
measurements in Fig.~\protect\ref{fig3}.} \label{fig7}
 \end{figure}

These results suggest us to use the transport properties to
investigate the exchange-bias effect, as a way to strengthen
the interpretation of the magnetization measurements done with the
SQUID. Similar to the SQUID measurements shown in Fig.~\ref{fig3},
Fig.~\ref{fig7} shows the MR and PHE measurements at two different
temperatures and after cooling the sample with a positive and
negative magnetic field of $\pm 1~$T. We observe that after the
first reversal process (arrows (1) and (1')) the MR shows a sharp
change, while in the second reversal process  (arrows (2) and
(2')) the change is more rounded and larger in amplitude compared
to the first one. This indicates that during the first
magnetization reversal there are less moments contributing to the
magnetization rotation compared to the subsequent reversal process
\cite{bre05}. The field shifting and asymmetries observed in the
MR and PHE are in agreement with the ones observed in the SQUID
measurements indicating that the same effects (domain wall
propagation and nucleation) contribute in a similar manner in
these properties. In both kinds of  measurements, we observe that at
the fields of, e.g. +0.5~T or -0.5~T, the initial and final values
are not the same, i.e. the MR and PHE show a clear irreversible
behavior after completing the field cycle. This irreversibility is
more evident in the PHE compared to the AMR results, and both
strengthen the existence of the $m_{\rm shift}$ effect measured
with the SQUID.

\begin{figure} %\vskip  9cm  \special{eps:  F:/PCTeXv4/figure1.EPS  x=8.2cm  y=6.2cm}
\includegraphics[width=1.1\columnwidth]{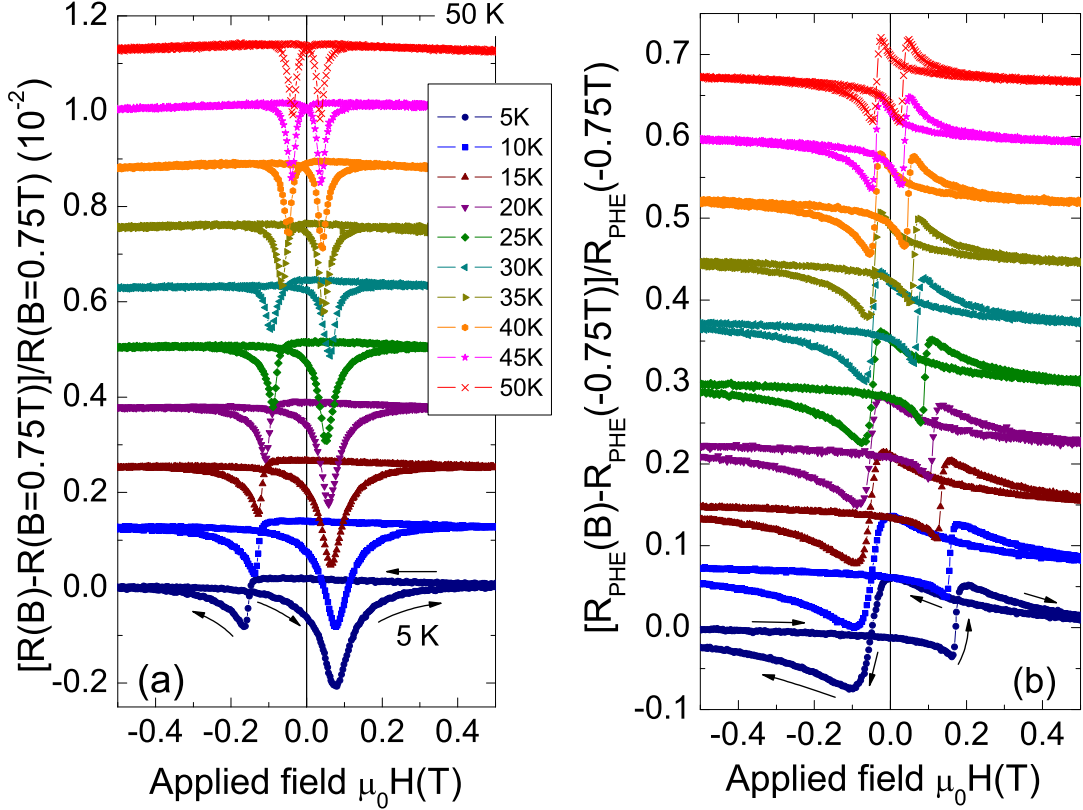}
\caption{Magnetoresistance (a) and the planar Hall effect (b)
measured using a positive and negative cooling field $|\mu_0
H_{\rm FC}| = 1~$T, respectively.} \label{fig8}
 \end{figure}

\begin{figure} %\vskip  9cm  \special{eps:  F:/PCTeXv4/figure1.EPS  x=8.2cm  y=6.2cm}
\includegraphics[width=0.7\columnwidth]{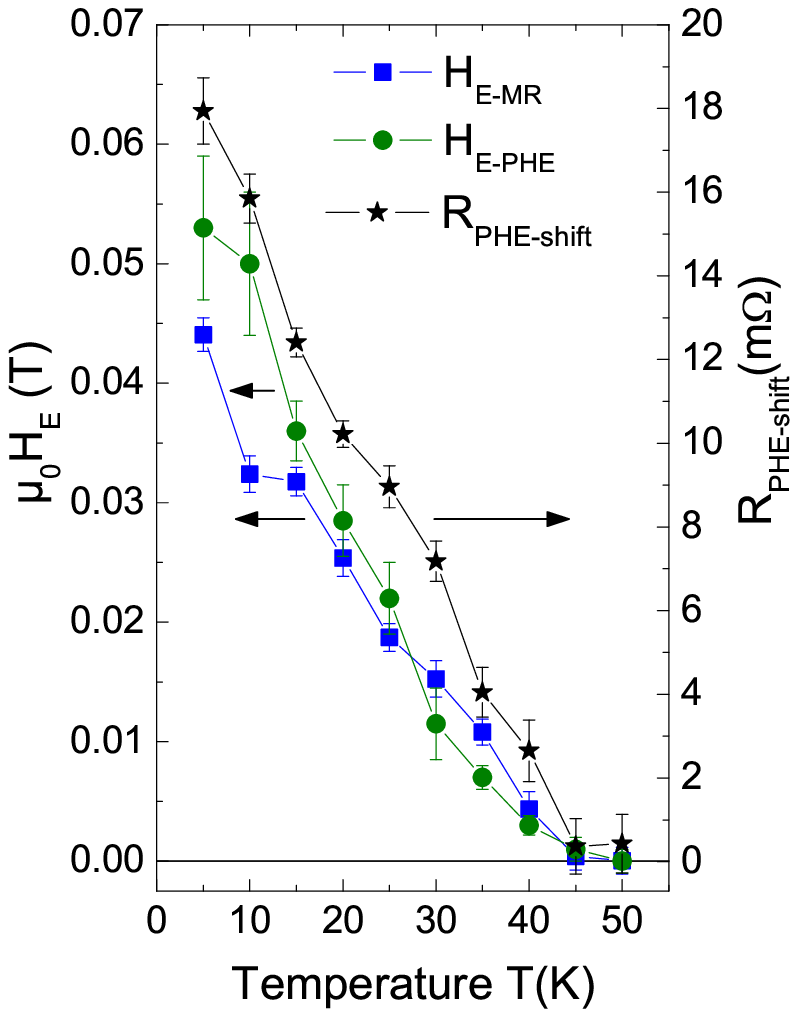}
\caption{Left $y-$axis: Exchange bias fields obtained from the
magnetoresistance ($H_{E-{\rm MR}}$) and from the planar Hall
effect ($H_{E-{\rm PHE}}$). Right $y-$axis: Resistance shift in
the PHE calculated at $\mu_0 H = -0.4~$T from the data in
Fig.~\protect\ref{fig8}(b).} \label{fig9}
 \end{figure}

Figure~\ref{fig8} shows the results obtained for the MR and the
PHE as a function of applied field at cooling fields $H_{\rm FC} =
\pm 1~$T at different constant temperatures. The exchange bias
effects are manifested in: - the field and saturation asymmetries,
- the different shape at the fields where the magnetization
direction changes, and - by the smearing of these characteristics
increasing temperature and their vanishing at or just above the
N\'eel temperature. Assuming that the two minima in the MR and the
sharp steps in the PHE define the coercive fields, we apply a
similar definition as before to estimate the exchange bias field
from these two properties $H_{E-{\rm MR}}$ and $H_{E-{\rm PHE}}$.
The temperature dependence of these exchange fields are shown in
Fig.~\ref{fig9}. The nearly linear behavior agrees with that
obtained from the SQUID measurements, see Fig.~\ref{fig5}. Note
that the absolute $H_E$ values obtained from the transport data
(Fig.~\ref{fig9}) are larger than for the SQUID data
(Fig.~\ref{fig5}) just because we compare data obtained at
different $H_{\rm FC}$.

The analogous to the magnetic moment shift observed in the SQUID
measurement appears also in the MR and PHR measurements, i.e. an
irreversibility in the resistance or Hall resistance at the same
field after completing a field cycle. As in the case of
magnetization measurements, we quantify this irreversibility by
the definition, in case of the PHE shift, $R_{{\rm PHE-shift}} =
R_{PHE}(-0.4$~T)$\uparrow - R_{PHE}( -0.4$~T)$\downarrow$, where
$R_{PHE}(-0.4$~T)$\uparrow$ means the planar Hall resistance at -0.4~T
increasing field from $H_{\rm FC} = -1~$T and
$R_H(-0.4$~T)$\downarrow$ the value obtained at the same field but
coming from positive fields. Figure~\ref{fig9} shows the obtained
temperature dependence of the parameter $R_{{\rm PHE-shift}}(T)$.
From the comparison with $H_E(T)$,  it is reasonable to assume
that the origin of the exchange bias parameter $R_{{\rm
PHE-shift}}(T)$ is the same as the $m_{\rm shift}$ from SQUID
measurements and should be directly correlated with the pinning of
domains or magnetic moments of the FM Co layer due to its common
interface with the AFM layer. \red{We stress that although the
measured transport properties provide information only of the Co
layer, it is clear that all exchange bias effects are related to
the influence of the AFM layer on the Co layer at the interface.}

Note that $R_{PHE-{\rm shift}}(T)$ does appear to show slight
changes of slope at 30~K and 20~K, which may be related to the
ferroelectric transition of the o-YMO layer, although they are not
as clear as for the $m_{\rm shift}(T)$ obtained from the SQUID
measurements. This difference between the temperature dependence
of these two quantities, $R_{PHE-{\rm shift}}(T)$ and $m_{\rm
shift}(T)$, is related to the different $H_{FC}$ used and in part
to the field we used to calculate $R_{PHE-{\rm shift}}(T)$. Note
that the $H_{FC}$ dependence of the $m_{\rm shift}$ is non
monotonous, i.e. $M_E \rightarrow 0$ for $H_{\rm FC} \rightarrow
\infty$ as well as at $H_{\rm FC} = 0$, as has been shown for
La$_{0.7}Sr_{0.3}$MnO$_3$/o-YMO bilayers \cite{zan11}.

\section{Magneto-transport and SQUID measurements of the Co/CoO wires}
\label{cocoo} 
\begin{figure} %\vskip  9cm  \special{eps:  F:/PCTeXv4/figure1.EPS  x=8.2cm  y=6.2cm}
\includegraphics[width=1.1\columnwidth]{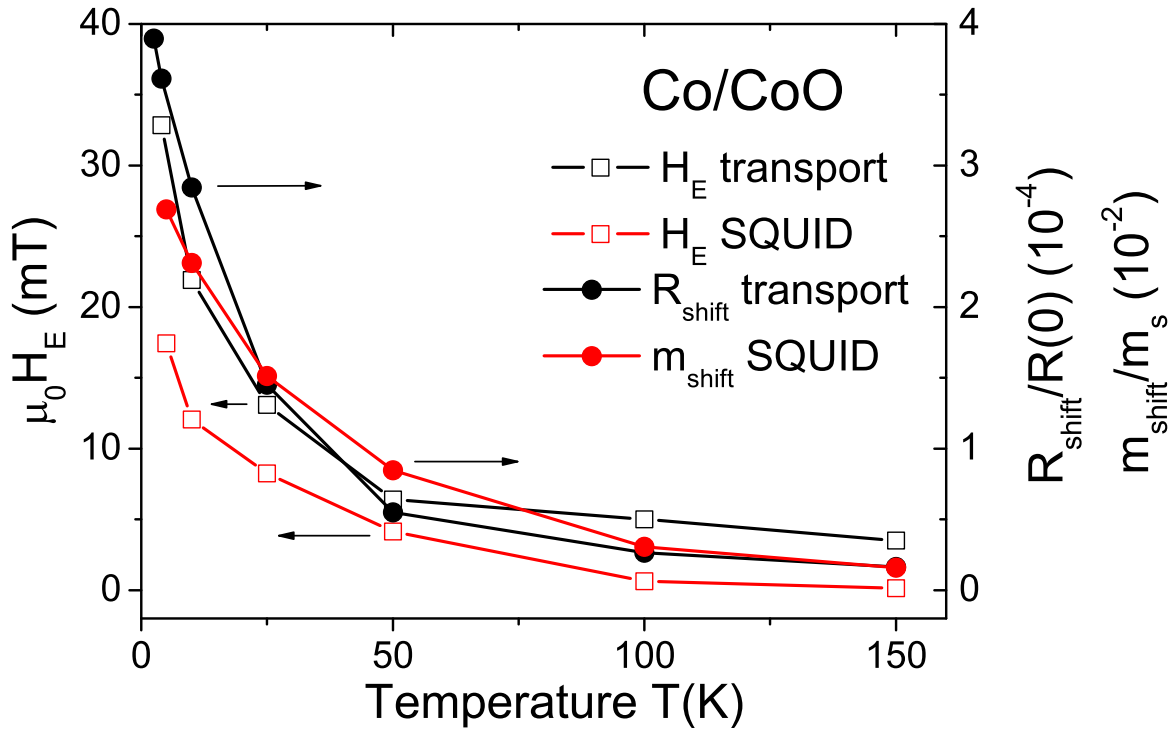}
\caption{Exchange bias field $H_E$ and normalized magnetic moment shift $m_{\rm shift}$
(right $y-$axis) measured with a SQUID on an array of 8,500 micro-wires of Co/CoO (red symbols). Similar
exchange bias parameters obtained from transport measurements (see Fig. 16) on
a single Co/CoO micro-wire.}
\label{fig16}
 \end{figure}

\red{ The exchange bias effects of the array of
8,500 micro-wires were measured with a SQUID.  For the FC measurements the sample
was mounted in the SQUID after the corresponding $H_{FC}$ was
applied. These measurements reveal shifts in both axis of the
hysteresis, similar to that shown in Fig.~\ref{fig3} for the
Co/YMO bilayer and we do not need to show them here.
Figure\,\ref{fig16} shows the temperature dependence of the
exchange bias field $H_E$ defined as before and the magnetization
shift ($M_E$-effect) normalized by its value at saturation. The
vertical magnetic moment shift $m_{\rm shift}$ is defined as
before and  at $\pm$0.1\,T. One can clearly recognize in
Fig.~\ref{fig16} that both exchange bias effects (red symbols) show a similar
temperature dependence. }

\begin{figure} %\vskip  9cm  \special{eps:  F:/PCTeXv4/figure1.EPS  x=8.2cm  y=6.2cm}
\includegraphics[width=1.1\columnwidth]{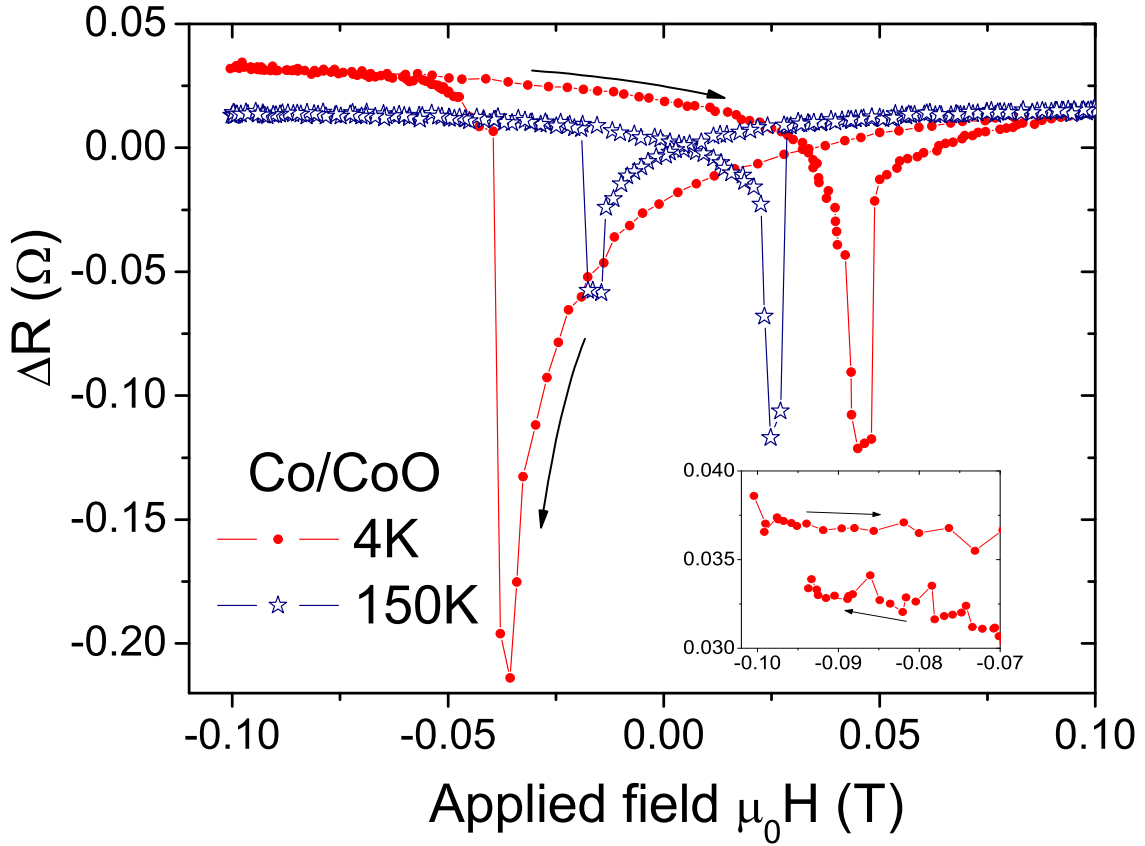}
\caption{Magnetoresistance defined as $\Delta R = R - R_{\rm
intersect}$ at two temperatures
 of a single Co/CoO micro-wire. $R_{\rm intersect}$ is defined as
 the resistance at which the increasing and decreasing field curves intersect.
 The sample was measured after cooling in a field of -8~T applied
 along the principal wire axis. The inset shows the measured curves at the first
field cycle near -0.1~T.} \label{fig15}
 \end{figure}

\red{A typical hysteresis in the magnetoresistance of a single
Co/CoO micro-wire is shown in Fig.~\ref{fig15} after cooling the
sample in a field of -8~T at two different temperatures. The
asymmetry in the coercive field (at the minima) as well as in the
saturation region $R_{\rm shift} = R(-0.1~$T$) - R(+0.1~$T)$ > 0$
can be clearly recognized in the figure at 4~K, similarly to the
one obtained for the bilayer Co/YMO, see Fig.~\ref{fig7}(a). The
red curve at 4K shown in the main panel is the hysteresis measured
the second time after one cycle to +0.1~T $\Rightarrow -0.1~$T. In
the inset we show the irreversibility in the resistance after the
first hysteresis, similar to the irreversibility measured in the
Co-YMO bilayer, see Fig.~\ref{fig7}(a). The two exchange bias
effects obtained from transport measurements of a single
micro-wire show similar temperature dependence between each other
and to the SQUID results, see Fig.~\ref{fig16}. In our samples the
exchange bias effects vanish  at about 150\,K. This may be due to
the small thickness of the AF CoO layer. A blocking temperature in
this range has been reported for equally thin layers~\cite{bun05}.
  We think that the results in Co/CoO micro-wires leave little doubt about the
  existence of both exchange bias effects. Because of the negligible conductance
  of the oxide layer in comparison with the Co part at the temperature of
  the measurements, the transport results also indicate that the
  origin of the magnetization shift comes from the
  AFM Co layer at the interface with the CoO layer, in agreement with the
  main results obtained for the Co/YMO bilayer.}

\section{Conclusion} In the present work we have systematically
investigated the exchange bias phenomenon in the novel Co/YMO
bilayer using SQUID magnetometry and magneto-transport properties.
The extracted exchange bias anomalies show the expected field
asymmetry, decreasing nearly linearly with temperature.
Additionally, a vertical shift is observed in all measurements,
which is related to the pinning of magnetic entities of the
ferromagnetic Co layer at the interface with the o-YMO layer. Both
exchange bias shifts vanish at or just above the N\'eel
temperature confirming also that these are related to the exchange
bias phenomenon.  From the magnetization values of Co and o-YMO
layers and the observed $m_{\rm shift}$ we can also conclude that
the measured $M_E$ effect in the bilayer is a direct contribution
from the FM layer. The magneto-transport results provide a clear
support to this conclusion. \red{The results obtained from similar
magnetization and transport measurements in Co/CoO micro-wires
support the main conclusions given above, and stress that the
observed phenomena are general and not because of any special
magnetic characteristics of the o-YMO layer.}

\ack{One of us (C.Z.) was supported  by  the S\"achsisches
Staatsministerium f\"ur Wissenschaft und Kunst under
4-7531.50-04-0361-09/1. This work was supported by the
Collaborative Research Center SFB 762 ``Functionality of Oxide
Interfaces"}

\section*{References}

%\bibliography{D:/DATA/Future_SFB/YMO-Co/bilayers,D:/DATA/hopg/magnetic_carbon} %office
%\bibliography{/Users/pablo/Documents/biblib/bilayers,/Users/pablo/Documents/biblib/magnetic_carbon} %casa

\end{document}